\documentclass[10pt,journal]{IEEEtran}

\makeatletter
\def\ps@headings{%
\def\@oddhead{\mbox{}\scriptsize\rightmark \hfil \thepage}%
\def\@evenhead{\scriptsize\thepage \hfil \leftmark\mbox{}}%
\def\@oddfoot{}%
\def\@evenfoot{}}

\makeatother

\ifCLASSOPTIONcompsoc

\else
\fi

\ifCLASSOPTIONcompsoc
  \ifCLASSOPTIONconference

  \else

  \fi
\else
\fi

\usepackage{amssymb,color,amsmath}
\usepackage[utf8]{inputenc}
\usepackage[english]{babel}
\usepackage{multirow,multicol}
\usepackage{amsthm}
\usepackage{cite,manfnt}
\usepackage[urlcolor=rltblue,colorlinks=true]{hyperref} %
\usepackage{arydshln}

\usepackage{subfigure}
\usepackage{paralist,color,clock}
\usepackage{graphicx}

\usepackage{algorithm}
\usepackage{algpseudocode}
\newcommand{\algrule}[1][.2pt]{\par\vskip.5\baselineskip\hrule height #1\par\vskip.5\baselineskip}

\usepackage{tikz}
\usetikzlibrary{fit,positioning}

\definecolor{rltblue}{rgb}{0,0,0.75}\usepackage[colorlinks=true,urlcolor=rltblue]{hyperref}

\newcommand{\nix}[1]{}

\begin{document}
\title{Towards an Efficient Voice Identification Using Wav2Vec2.0 and HuBERT Based on the Quran Reciters Dataset}

\author{Aly Moustafa$^\dag$ ~~~~~~~~~~~Salah A. Aly$^\ddag$\thanks{---------------------------------- \\ $^\dag$Faculty of Computers and Artificial Intelligence,  Helwan  University, Egypt\\
$^\ddag$CS and Math section,   Faculty of Science,  Fayoum University, Egypt\\ }\\
}

\date{}
\maketitle              

\begin{abstract}
Current authentication and trusted systems depend on classical and biometric methods to recognize or authorize users.  Such methods include audio speech recognitions, eye, and finger signatures.  Recent tools utilize deep learning and transformers to achieve better results. In this paper, we develop a deep learning constructed model for Arabic speakers’ identification by using Wav2Vec2.0 and HuBERT audio representation learning tools. The end-to-end Wav2Vec2.0 paradigm acquires contextualized speech representations learning's by randomly masking a set of feature vectors, and then applies a transformer neural network. We employ an MLP classifier that is able to differentiate between invariant labeled classes. We show several experimental results that safeguard the high accuracy of the proposed model. The experiments ensure that an arbitrary wave signal for a certain speaker can be identified with $98\%$ and $97.1\%$ accuracies in the cases of Wav2Vec2.0 and HuBERT, respectively.

\end{abstract}

\section{Introduction}

With the huge widespread of social media and audio networks, it is mandatory to develop machine learning tools to identify and recognize joined connected speakers in joint conversations.  Such tools are able to detect fake audio recordings and recognize falsehood human speeches.  Communication systems over various social media need to first authenticate communicating users before actual conversations are being congregated.

The sounds other than the speech of humans  (objects) are defined as background noise. Background noise can affect some attributes of speeches: intelligibility, clearness, and quality~\cite{Michelsanti2019}.

The problem of speech  identification based on a given set of speeches has been investigated recently by many researchers~\cite{Bai2012},~\cite{Shahin2021a},~ \cite{Shahin2021b},~\cite{Asda2016}. The goal is to produce a speech-to-identify and recognize machine learning system. Speaker identification can be identified based on either text-independent or text-dependant.

Recent work on speech recognition focuses on the way speakers are stressed, emotional, and disguised in their speeches~\cite{aly2021a}. In this work, we aim to develop a deep learning model for voice identification in Arabic speech.  We use a Quranic dataset developed by ~\cite{Nahar2019}.

Unlike Google speech recognition API~\cite{GoogleSTT}, Wav2vec2.0, with its remarkable series wav2vec, wav2vecU Unsupervised~\cite{Baevski2021wav2vecU},  is a well-recognized API library recently developed by Facebook folks~\cite{wav2vec2}. Wav2Vec2.0 is self-supervised learning of representations from raw audio data, that assured the framework can enable automatic speech recognition with just only 10 minutes of transcribed audio data.  Wav2Vec2.0 model is pre-trained on 16 kHz frequency. The mentioned model wav2vec2.0 enables speech recognition frameworks at a word error rate (WER) of $8.6$ percent on noisy speech and $5.2$ percent on clean speech on the standard LibriSpeech benchmark~\cite{wav2vec2}. A cross-lingual approach model, called XLSR,  can learn speech units common to several languages. Wav2vec2.0 was trained by predicting speech units for masked parts of the audio, in the same manner as Bidirectional Encoder Representations from Transformers (BERT).

Recently, wav2vecU Unsupervised model is proposed to train speech recognition models without any labeled data~~\cite{Baevski2021wav2vecU}. Wav2vec-U decreases the phoneme error rate from 26.1 to 11.3 on the TIMIT standard in comprising to unsupervised wav2vec.

The paper structure is described as follows.  In Section~\ref{sec:relatedwork} we present related work. In Sections~\ref{sec:dataset} we describe the AR-DAD dataset that is used in our methods, which is developed in Section~\ref{sec:methods}. In Section~\ref{sec:analysis} we show some analysis for the deep learning algorithms, and simulation studies for the proposed algorithms are demonstrated in  Section~\ref{sec:simulation}. Finally, the paper is concluded in Section~\ref{sec:conclusion}.

\begin{figure*}[t]
  \centering
  \includegraphics[width=\textwidth,height=7cm]{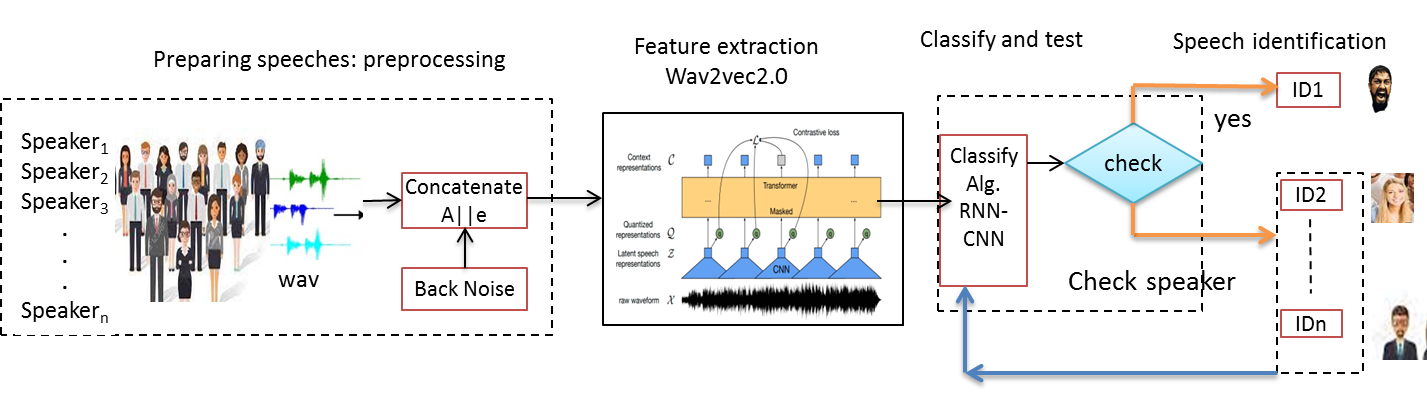}

  \caption{Architecture of the system for identifying a set of speeches}\label{fig:model1}
\end{figure*}

\section{Related Work}\label{sec:relatedwork}

There have been several research efforts in speaker identifications. Several search attempts for the Quran reciter’s identification and verification are proposed, using recurrent neural networks and deep learning, see~\cite{Ahmed2017,Sajjad2017,Lataifeha2020,Nahar2019,Elnagar2018,Lataifeha2020}.

Recently, Lataifeha etc.~\cite{Lataifeha2020} introduced a speaker identification model of 30 known reciters of the Holy Quran.  In their work, an audio assembled dataset is presented, which consists of 15810 audio clips that are used for training, and another 397 audio clips that are used for top imitators.  In addition, for classification, six top-achieving classical and two deep learning classiﬁers are used.

Noteworthy, Elnagar etc.~\cite{Elnagar2020} described a comparative analysis for a supervised classification system of Quranic audio clips of several reciters and evaluated different classifiers' performance to assess the model.  Elnagar etc.~\cite{Elnagar2018} described a supervised classification system of the holy Quran audio clips of various  reciters and constructed a representative dataset of audio clips for seven popular reciters, particularly from Saudi Arabia.

Sajjad etc.~\cite{Sajjad2017} introduced a model for Speaker Identiﬁcation \& Veriﬁcation Using MFCC \& SVM.

Khelifa etc.~\cite{Khelifa2017StrategiesFI} demonstrated a practical model for  an optimal ASR system development  for Quranic audio
recognition, in which a statistical approach of Hidden Markov Models (HMMs)  and the Cambridge HTK tools are used.

Mohamed etc.~\cite{Mohamed2014VirtualLS} developed a virtual learning recitations system for Sighted and Blind Students and developed an efficient speech recognition engine that is a speaker and accent independent.

The authors in~\cite{Shahin2021a} proposed a classifier that is based on exploiting supervised Convolutional Neural Network (CNN) using three distinct speech databases:  local Arabic Emirati-accented, global English SUSAS, and global English RAVDESS.  The most recent used classifiers are the Gaussian Mixture Model, Vector Quantization, and  Hidden Markov Model. The author in this work developed a new classifier based on CNN model in stressful talking environments using English and Arabic Emirati-accented databases~\cite{Shahin2021a}.

\section{AR-DAD Dataset}\label{sec:dataset}

This work uses Arabic Diversified Audio Dataset~\cite{ARDADdataset}, denoted AR-DAD, that is delivered online by several reciters from several countries.  AR-DAD ~\cite{Lataifeh2020ArDADAD,ARDADdataset} includes approximately 15810  Arabic-based audio clips, which are taken from 37 chapters from the Holy Quran for 30 well-known reciters.   In addition, there are additional 397 audio clips for 12 competent imitators of those distinguished reciters that have the most downloaded statistics and are authorized by many agencies.

The AR-DAD dataset is prearranged into three main directories: reciters directory, Textual directory, and Imitators’ directory. In this work, we only used 1000 records of the AR-DAD dataset with only used 10 reciters.

\section{Proposed Methods}\label{sec:methods}
In this section, we present the feature extraction and classification.  In our model, we acquire raw audio clips from 10 reciters of the Holy Quran.

\subsection{Feature Extraction}

In our work, we use two models for Feature extractions: wav2vec2.0 and HuBERT. The reason we use these models is that they can obtain high-level contextual representation and learn basic units for less labeled data. It can also capture aspects without a clear segmentation for words.

\textbf{Wav2vec2.0~\cite{wav2vec2}:}
wav2vec2.0 process the raw audio data with a multilayer convolutional neural network  (CNN) to obtain latent audio representations of 25ms each. The representations are encapsulated into a quantizer and a transformer for feature extraction and selection as shown in Fig.~\ref{fig:model1}.
The quantizer and transformer processes are described in detail in~\cite{wav2vec2}.  Quantization is achieved via Gumbel and K-means. Wav2vec2.0  properties include fully convolutional, binary cross-entropy loss, and its representations used to improve ASR tasks.
Codebook diversity penalty to encourage more codes to be used. The strength of wav2vec2.0 comes from the idea that it performs joint learning between quantization of the latent representation. This feature was not existing in previous versions of  Wav2vec.

\textbf{HuBERT:}
 Wei-Ning Hsu etc. proposed  Hidden-Unit BERT (HuBERT)~\cite{hsu2021hubert}; Self-Supervised Speech Representation Learning by Masked Prediction of Hidden Units in June 2021 as a remarkable tool for Self-supervised approaches for speech representation training and learning. As stated in the seminal paper, HuBERT employs the prediction loss above the masked regions only that influences the model to learn a combined acoustic and contextual model over the continuous inputs. It relies mainly on the consistency of the unsupervised clustering step rather than the intrinsic quality of the allocated cluster labels.
 In a performance study, the HuBERT either matches or improves over wav2vec 2.0 performance on the Librispeech (960h) and Libri-light (60,000h) benchmarks with 10min, 1h, 10h, 100h, and 960h fine-tuning subsets for a simple k-means teacher of 100 clusters, and using only two iterations of clustering, see~\cite{hsu2021hubert}. To produce a better representation, HuBERT replies on the consistency of unsupervised learning using k-means teacher.

\subsection{Classifiers}

Connectionist Temporal Classification (CTC) is a sequence modeling algorithm that is plugged in wav2vec2.0 overcomes the problem of accurate alignment and variation lengths of inputs and outputs. CTC is suitable for speech and handwriting recognition.

We have 10 classes and our target is to identify the reciters' voices. We perform the classification process by using three different algorithms: MLP, RNN, and CNN.  We perform end-to-end training in which we added a classification head in the wave2vec and HuBERT.

\begin{table}
  \centering
  \caption{Average accuracy of various classifiers: Speaker identification accuracy evaluation using based on CNN, NB, RBF, SVM, KNN, andMLP}\label{classify_table}
\begin{tabular}{|c|c|}
  \hline
  classifier & Accuracy\\
  \hline
  MLP &  0000\\
  CNN & 0000 \\
  Bi-lstm& 0000 \\

  \hline
\end{tabular}
\end{table}

As stated in Table~\ref{classify_table}, different classifiers achieve the greatest results under neutral giving conditions only without any stressful conditions. Moreover, the average of each classifier has been calculated and the result shows that CNN is superior to each of MLP, KNN, NB, RBF, and SVM in all talking conditions.

\section{Training Procedure}\label{sec:analysis}

We train the HuBERT and wav2vec2.0 models for 1000 records with 52 min. In detail, HuBERT-base and HuBERT-large took 240 steps and a long time to converge. This is due to the fact that HuBERT is trained in multiple languages, and did not fine-tune previously in Arabic speeches. In contrast, there is a version of wav2vec2.0 large, which is fine-tuned in Arabic speeches, and it makes the convergence fast.

The algorithms steps can be stated as follows.
Sample a mini-batch of feature vector lab pairs from the training set.
Compute $h_o$ as the initialization of the attention-network layer (A.N.L.).
Compute loss between predicted label $y$ and target $Y$

Let $TL$ and $VL$ be the training loss and validation loss, respectively.
\begin{algorithm}
\caption{Training Procedure of AR-Wav2Wav speaker identification Algorithm}\label{alg:cap}
  \hspace*{\algorithmicindent} \textbf{Input:} {Raw audio sequence A} \\
 \hspace*{\algorithmicindent}  \textbf{Output: }{ Acoustic Speaker Identification}
  \algrule
  \begin{algorithmic}[1]
    \Procedure{AR-wave2wav}{$A$}\Comment{Audio Identification}
    \State ${A}\gets []$
    \State $V \gets []$

      \vskip 5mm
      \For{i =1 to \text{max-iter}}
    \State Sample a mini-batch of  pairs from ${A}$
    \State \texttt{ $V \gets wav2vec~FeatureExtractor({A_i})$}
    \State Compute $h_o$ as the init. of    A.N.L.
    \State Compute loss for predicted  $y$ and target $Y$
    \State \textbf{Print }   $TL$
    \State \textbf{Print } $VL$
    \EndFor
    \State Print ${a_j}$ speaker identification
    \EndProcedure
  \end{algorithmic}
\end{algorithm}

\begin{figure*}[t]
  \centering
  \includegraphics[width=5.9cm,height=6cm]{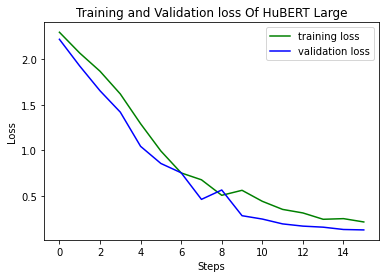}
  \includegraphics[width=6cm,height=6cm]{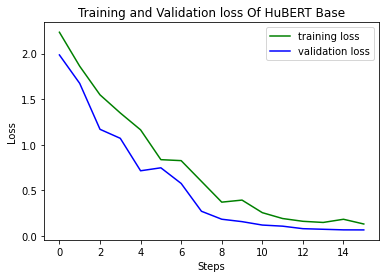}
  \includegraphics[width=5.9cm,height=6cm]{fig/train_val_loss_hubert_large.png}
  \caption{Results and accuracy of the proposed models}\label{fig:f1score}
\end{figure*}

We conducted weighted cross-entropy to overcome the problem of the unbalanced dataset.

\section{Results and Performance Evaluations}\label{sec:simulation}

We performed several experimental tests to measure the success of our proposed model. The measurement accuracy is described in Table~\ref{table:result1}.

          \begin{table}[h]
            \centering
             \caption{Measured accuracy of the three models}\label{table:result1}
        \begin{tabular}{|c|c|c|c|c|c|}
        \hline
          Data size & Length & Model   & Clip &reciters & F1-\\
         Clips &&&Len.&&Score\\
          \hline
          1000  & 53-Min & Wav2vec.2.0 Large     & $<=$ 4 & 10&  97\% \\
          1000  & 53-Min & HuBERT Base & $<= 4$  & 10& 98\% \\
          1000  & 53-Min & HuBERT Large & $<= 4$ & 10& 99\% \\
          \hline
        \end{tabular}
        \end{table}

\subsection{F-1 Score}

To measure the precision and recall, we used F1-score as described in~\ref{fig:f1score}.

\begin{equation}\label{}
  F_1= 2 \frac{precison. recall}{precision + recall}
\end{equation}

The reason we use the F-1 score is that it gives better measurement for unbalanced data.

Figure~\ref{fig:f1score} shows the validation accuracy across the three models: wav2vec2.0, HuBERT large, and HuBERT base.  At the beginning of the training, Wav2vec2.0 converges much faster than other models because it was previously fine-tuned on Arabic corpus.  However, after completion of the training process,  HuBERT large and HuBERT achieved better validation accuracy.

\subsection{Validation Loss and Training Loss}

\begin{figure}[h]
  \centering
  \includegraphics[width=8.5cm,height=7cm]{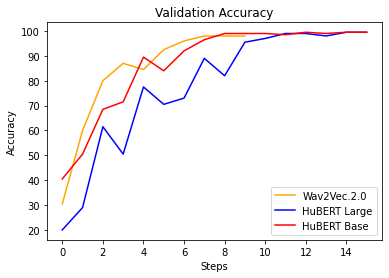}
  \caption{Results and accuracy of the proposed  models}\label{fig:accuracy_all_models}
\end{figure}

Figure~~\ref{fig:accuracy_all_models} presents a comparison between the accuracy of wav2vec2.0, HuBERT large, and HuBERT base models.

\begin{figure}[h]
  \centering
  \includegraphics[width=8.5cm,height=7cm]{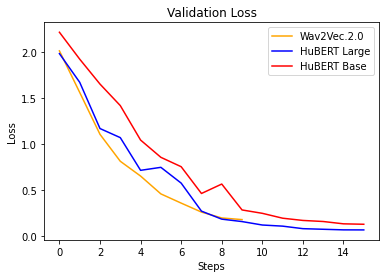}
  \caption{Results and accuracy of the validation loss all models}\label{fig:validation_loss_all_models}
\end{figure}

Figure~~\ref{fig:validation_loss_all_models} presents the comparison between training and validation loss of wav2vec2.0, HuBERT large, and HuBERT base models. After certain steps, the training loss converges to validation loss.

\begin{figure*}[t]
  \centering
  \includegraphics[width=5.8cm,height=6cm]{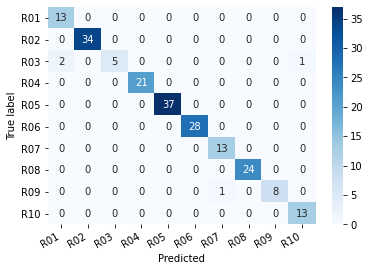}
   \includegraphics[width=6cm,height=6cm]{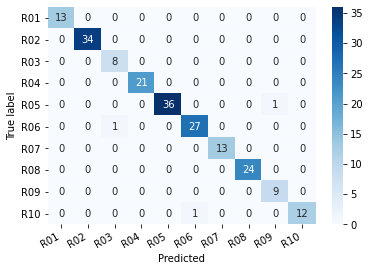}
  \includegraphics[width=5.8cm,height=6cm]{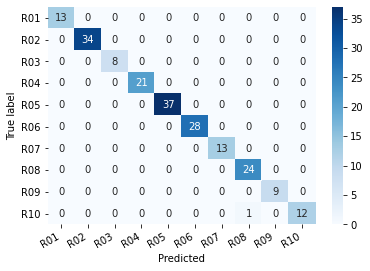}
  \caption{(a) Results and accuracy of the wav2vec2.0 conf matrix, (b) Results and accuracy of the HuBERT base conf matrix, (c) Results and accuracy of the HuBERT large conf matrix}\label{fig:all_confusion_matrix}
\end{figure*}
\subsection{Confusion Matrix}

Figure~\ref{fig:all_confusion_matrix} displays that the prediction matrix of the wa2vec confusion matrix model among all ten speakers. The figure shows that speaker number five has the highest prediction value. Figure~\ref{fig:all_confusion_matrix} (a) shows that R01 has an intersection with R03, and the wav2vec2.0 model was not able to differentiate between them.  Also, in Figure 6 (b) the HuBERT base model was not able to differentiate between R06 and R10. These wav2vec2, HuBERT base, HuBERT large models miss only $4$, $3$, and $1$ miss-classified samples, respectively. It was noticed that the three models did not miss in the same sample.

\section{Discussions and Future Research Directions}

The titanic success of neural networks in the machine and deep learning are transforming the interaction between humans and objects around in the globe. The famous types of neural networks including CNN, RNN, ANN are at the core of the machine and deep learning revolution with various real applications specifically in speech recognition, image, and video transformations.

In this work, we used only $1000$ records. Previous conducted by~\cite{Lataifeha2020} used $15000$ records and obtained $98\%$.

Speaker speech identification has various applications. Some recent trusted systems depend on voice recognition in order to access the system.

Other speaker identification tools used classical neural networks such as RNN and CNN to be able to recognize the target speaker.  New tools for Automated Speech Recognition (ASR) did achieve high accuracy for two reasons: \begin{itemize} \item recently proposed speech recognitions representation tools such as wav2vec2.0 and HuBERT present high accuracy and low validation loss. \item applying innovative approaches that use deep learning concepts. \item several datasets have been available online that made is easy for researchers to download, then train and test their developed models.
\end{itemize}

As a research direction, HuBERT can be fine-turned for a large Arabic corpus. This may improve the performance of the model and obtain a better representation of the Arabic speeches.

A large dataset can also be tested using this model to classify more audio acoustics.

\section{Conclusion}\label{sec:conclusion}

We developed wav2wav2.0 and HuBERT deep learning models for Arabic speakers' identification. We implemented these learning models and demonstrated their results on the Arabic audio BASED dataset. Several experiments are performed using wav2vec2 and HuBERT different validation techniques.  The model is successfully performed by using wav2vec2 and yielded an accuracy of $96.23\%$ with data augmentation. In future work, we plan to extend the proposed method to incorporate more feature sets and increase the size of the dataset for words, sentences, paragraph recognition.

The proposed model can be deployed in various practical systems in the cases of biometric gate entrance, bank account checking, critical security access control, etc.
\bigskip

\section*{Acknowledgement}
This research is partially funded by a grant from the academy of scientific research and technology (ASRT), 2020-2021, research grant number 6547.

\bigskip

\bibliographystyle{plain} 

\begin{thebibliography}{}

\end{thebibliography}


\begin{thebibliography}{20}

\bibitem{Michelsanti2019}
D.~Michelsanti, S.~S. Zheng-Hua~Tana, and J.~Jensen, ``Deep-learning-based
  audio-visual speech enhancement in presence of lombard effect,'' {\em Speech
  Communication, Elsevier BV, arXiv:1905.12605v1}, vol.~115, p.~38–50, 2019.

\bibitem{Bai2012}
Z.~Bai and X.-L. Zhang, ``Speaker recognition based on deep learning an
  overview,'' {\em Neural Networks arXiv:2012.00931v2 [eess.AS] 4 Apr 2021},
  vol.~140, p.~65–99, 2021.

\bibitem{Shahin2021a}
I.~Shahin, A.~B. Nassif, and N.~A.~A. Hindawi, ``Speaker identification in
  stressful talking environments based on convolutional neural network,'' {\em
  International Journal of Speech Technology}, July 2021.

\bibitem{Shahin2021b}
smail Shahin, A.~B. Nassif, N.~Nemmour, A.~Elnagar, A.~Alhudhaif, and K.~Polat,
  ``Novel hybrid dnn approaches for speaker verification in emotional and
  stressful talking environments,'' {\em Neural Computing and Applications},
  June 2021.

\bibitem{Asda2016}
T.~M.~H. Asda, T.~S. Gunawan, M.~Kartiwi, and H.~Mansor, ``Development of quran
  reciter identiﬁcation system using mfcc and neural network,'' {\em
  TELKOMNIKA Indonesian Journal of Electrical Engineering}, vol.~17 (1,
  p.~168–175, 2016.
\newblock doi:10.11591/ijeecs.v1.i1.pp168-175.

\bibitem{aly2021a}
O.~Mohamed and S.~A. Aly, ``{ASER}: Arabic speech emotion recognition employing
  {Wav2vec2.0} and {HuBERT} based on {BAVED} dataset,'' {\em Transactions on
  Machine Learning and Artificial Intelligence {(TMLAI}) Journal},
  arXiv:2110.04425, October, 2021.

\bibitem{Nahar2019}
K.~M.~O. Nahar, M.~Al-Shannaq, A.~Manasrah, R.~Alshorman, and I.~Alazzam, ``A
  holy quran reader/reciter identiﬁcation system using support vector
  machine,'' {\em International Journal of Machine Learning and Computing},
  vol.~9, no.~4, p.~458–464, 2019.

\bibitem{GoogleSTT}
G.~C. https://cloud.google.com/speech-to text.

\bibitem{Baevski2021wav2vecU}
A.~Baevski, W.-N. Hsu, A.~Conneau, and M.~Auli, ``Unsupervised speech
  recognition,'' {\em arXiv:2105.11084v1}, 2021.

\bibitem{wav2vec2}
``{Facebook {Wav2Vec2.0}:
  https://ai.facebook.com/blog/wav2vec-20-learning-the-structure-of-speech-from-raw-audio/},''

\bibitem{Ahmed2017}
A.~H. Ahmed and S.~M. Abdo, ``Veriﬁcation system for quran recitation
  recordings,'' {\em International Journal of Computer Applications},
  vol.~doi:10.5120/ijca2017913493, p.~6–11, 163 (4) (2017).

\bibitem{Sajjad2017}
A.~Sajjad, A.~Shirazi, N.~Tabassum, M.~Saquib, and N.~Sheikh, ``Speaker
  identiﬁcation \& veriﬁcation using mfcc \& svm,'' {\em International
  Research Journal of Engineering and Technology(IRJET) 4 (2)}, p.~1950–1953,
  2017.

\bibitem{Lataifeha2020}
M.~Lataifeh, A.~Elnagar, I.~Shahin, and A.~B. Nassif, ``Arabic audio clips:
  Identiﬁcation and discrimination of authentic cantillations from
  imitations,'' {\em Neurocomputing}, September 2020.

\bibitem{Elnagar2018}
A.~Elnagar, R.~Ismail, B.~Alattas, and A.~Alfalasi, ``Automatic classiﬁcation
  of reciters of quranic audio clips,'' {\em 2018 IEEE/ACS 15th International
  Conference on Computer Systems and Applications (AICCSA)}, p.~1–6, 2018.

\bibitem{Elnagar2020}
A.~Elnagar and M.~Lataifeh, ``Predicting quranic audio clips reciters using
  classical machine learning algorithms: A comparative study,'' {\em Computer
  Science}, 2020.

\bibitem{Khelifa2017StrategiesFI}
M.~O.~M. Khelifa, M.~Belkasmi, Y.~Elhadj, and Y.~Abdellah, ``Strategies for
  implementing an optimal asr system for quranic recitation recognition,'' {\em
  International Journal of Computer Applications}, vol.~172, pp.~35--41, 2017.

\bibitem{Mohamed2014VirtualLS}
S.~Mohamed, A.~S. Hassanin, and M.~T.~B. Othman, ``Virtual learning system
  (miqra’ah) for quran recitations for sighted and blind students,'' {\em
  Journal of Software Engineering and Applications}, vol.~07, pp.~195--205,
  2014.

\bibitem{ARDADdataset}
I.~endowment site for Holy Quran~recitations: http://www.everyayah.com

\bibitem{Lataifeh2020ArDADAD}
M.~Lataifeh and A.~Elnagar, ``Ar-dad: Arabic diversified audio dataset,'' {\em
  Data in Brief}, vol.~33, 2020.

\bibitem{hsu2021hubert}
W.-N. Hsu, B.~Bolte, Y.-H.~H. Tsai, K.~Lakhotia, R.~Salakhutdinov, and
  A.~Mohamed, ``{HuBERT}: Self-supervised speech representation learning by
  masked prediction of hidden units,'' preprint arXiv:2106.07447, 2021.

\end{thebibliography}


\end{document}